\documentstyle[twoside]{article}

\catcode`\@=11
\long\def\@makefntext#1{
\protect\noindent \hbox to 3.2pt {\hskip-.9pt  
$^{{\eightrm\@thefnmark}}$\hfil}#1\hfill}		

\def\@makefnmark{\hbox to 0pt{$^{\@thefnmark}$\hss}}	
	
\def\ps@myheadings{\let\@mkboth\@gobbletwo
\def\@oddhead{\hbox{}
\rightmark\hfil\eightrm\thepage}   
\def\@oddfoot{}\def\@evenhead{\eightrm\thepage\hfil
\leftmark\hbox{}}\def\@evenfoot{}
\def\sectionmark##1{}\def\subsectionmark##1{}}

\oddsidemargin=\evensidemargin
\newcounter{sectionc}\newcounter{subsectionc}\newcounter{subsubsectionc}
\renewcommand{\section}[1] {\vspace{12pt}\addtocounter{sectionc}{1} 
\setcounter{subsectionc}{0}\setcounter{subsubsectionc}{0}\noindent 
	{\tenbf\thesectionc. #1}\par\vspace{5pt}}
\renewcommand{\subsection}[1] {\vspace{12pt}\addtocounter{subsectionc}{1} 
	\setcounter{subsubsectionc}{0}\noindent 
	{\bf\thesectionc.\thesubsectionc. {\kern1pt \bfit #1}}\par\vspace{5pt}}
\renewcommand{\subsubsection}[1] {\vspace{12pt}\addtocounter{subsubsectionc}{1}
	\noindent{\tenrm\thesectionc.\thesubsectionc.\thesubsubsectionc.
	{\kern1pt \tenit #1}}\par\vspace{5pt}}
\newcommand{\nonumsection}[1] {\vspace{12pt}\noindent{\tenbf #1}
	\par\vspace{5pt}}

\newcounter{appendixc}
\newcounter{subappendixc}[appendixc]
\newcounter{subsubappendixc}[subappendixc]
\renewcommand{\thesubappendixc}{\Alph{appendixc}.\arabic{subappendixc}}
\renewcommand{\thesubsubappendixc}
	{\Alph{appendixc}.\arabic{subappendixc}.\arabic{subsubappendixc}}

\renewcommand{\appendix}[1] {\vspace{12pt}
        \refstepcounter{appendixc}
        \setcounter{figure}{0}
        \setcounter{table}{0}
        \setcounter{lemma}{0}
        \setcounter{theorem}{0}
        \setcounter{corollary}{0}
        \setcounter{definition}{0}
        \setcounter{equation}{0}
        \renewcommand{\thefigure}{\Alph{appendixc}.\arabic{figure}}
        \renewcommand{\thetable}{\Alph{appendixc}.\arabic{table}}
        \renewcommand{\theappendixc}{\Alph{appendixc}}
        \renewcommand{\thelemma}{\Alph{appendixc}.\arabic{lemma}}
        \renewcommand{\thetheorem}{\Alph{appendixc}.\arabic{theorem}}
        \renewcommand{\thedefinition}{\Alph{appendixc}.\arabic{definition}}
        \renewcommand{\thecorollary}{\Alph{appendixc}.\arabic{corollary}}
        \renewcommand{\theequation}{\Alph{appendixc}.\arabic{equation}}
        \noindent{\tenbf Appendix \theappendixc #1}\par\vspace{5pt}}
\newcommand{\subappendix}[1] {\vspace{12pt}
        \refstepcounter{subappendixc}
        \noindent{\bf Appendix \thesubappendixc. {\kern1pt \bfit #1}}
	\par\vspace{5pt}}
\newcommand{\subsubappendix}[1] {\vspace{12pt}
        \refstepcounter{subsubappendixc}
        \noindent{\rm Appendix \thesubsubappendixc. {\kern1pt \tenit #1}}
	\par\vspace{5pt}}

\newcommand{\textlineskip}{\baselineskip=13pt}
\newcommand{\smalllineskip}{\baselineskip=10pt}

\def\eightcirc{
\begin{picture}(0,0)
\put(4.4,1.8){\circle{6.5}}
\end{picture}}
\def\eightcopyright{\eightcirc\kern2.7pt\hbox{\eightrm c}} 

\newcommand{\copyrightheading}[1]
	{\vspace*{-2.5cm}\smalllineskip{\flushleft
	{\footnotesize International Journal of Modern Physics A, #1}\\
	{\footnotesize $\eightcopyright$\, World Scientific Publishing
	 Company}\\
	 }}

\def\abstracts#1#2#3{{
	\centering{\begin{minipage}{4.5in}\baselineskip=10pt\footnotesize
	\parindent=0pt #1\par 
	\parindent=15pt #2\par
	\parindent=15pt #3
	\end{minipage}}\par}} 

\newcommand{\bibit}{\nineit}

\renewenvironment{thebibliography}[1]
	{\frenchspacing
	 \ninerm\baselineskip=11pt
	 \begin{list}{\arabic{enumi}.}
	{\usecounter{enumi}\setlength{\parsep}{0pt}
	 \setlength{\leftmargin 12.7pt}{\rightmargin 0pt} 
	 \setlength{\itemsep}{0pt} \settowidth
	{\labelwidth}{#1.}\sloppy}}{\end{list}}

\newcounter{itemlistc}
\newcounter{romanlistc}
\newcounter{alphlistc}
\newcounter{arabiclistc}

\newenvironment{romanlist}
	{\setcounter{romanlistc}{0}
	 \begin{list}{$($\roman{romanlistc}$)$}
	{\usecounter{romanlistc}
	 \setlength{\parsep}{0pt}
	 \setlength{\itemsep}{0pt}}}{\end{list}}

\newcommand{\fcaption}[1]{
        \refstepcounter{figure}
        \setbox\@tempboxa = \hbox{\footnotesize Fig.~\thefigure. #1}
        \ifdim \wd\@tempboxa > 5in
           {\begin{center}
        \parbox{5in}{\footnotesize\smalllineskip Fig.~\thefigure. #1}
            \end{center}}
        \else
             {\begin{center}
             {\footnotesize Fig.~\thefigure. #1}
              \end{center}}
        \fi}

\newcommand{\tcaption}[1]{
        \refstepcounter{table}
        \setbox\@tempboxa = \hbox{\footnotesize Table~\thetable. #1}
        \ifdim \wd\@tempboxa > 5in
           {\begin{center}
        \parbox{5in}{\footnotesize\smalllineskip Table~\thetable. #1}
            \end{center}}
        \else
             {\begin{center}
             {\footnotesize Table~\thetable. #1}
              \end{center}}
        \fi}

\def\@citex[#1]#2{\if@filesw\immediate\write\@auxout
	{\string\citation{#2}}\fi
\def\@citea{}\@cite{\@for\@citeb:=#2\do
	{\@citea\def\@citea{,}\@ifundefined
	{b@\@citeb}{{\bf ?}\@warning
	{Citation `\@citeb' on page \thepage \space undefined}}
	{\csname b@\@citeb\endcsname}}}{#1}}

\newif\if@cghi
\def\cite{\@cghitrue\@ifnextchar [{\@tempswatrue
	\@citex}{\@tempswafalse\@citex[]}}
\def\citelow{\@cghifalse\@ifnextchar [{\@tempswatrue
	\@citex}{\@tempswafalse\@citex[]}}
\def\@cite#1#2{{$\null^{#1}$\if@tempswa\typeout
	{IJCGA warning: optional citation argument 
	ignored: `#2'} \fi}}

\def\pmb#1{\setbox0=\hbox{#1}
	\kern-.025em\copy0\kern-\wd0
	\kern.05em\copy0\kern-\wd0
	\kern-.025em\raise.0433em\box0}

\def\fnt#1#2{\footnotetext{\kern-.3em
	{$^{\mbox{\scriptsize #1}}$}{#2}}}

\def\fpage#1{\begingroup
\voffset=.3in
\thispagestyle{empty}\begin{table}[b]\centerline{\footnotesize #1}
	\end{table}\endgroup}

\def\runninghead#1#2{\pagestyle{myheadings}
\markboth{{\protect\footnotesize\it{\quad #1}}\hfill}
{\hfill{\protect\footnotesize\it{#2\quad}}}}
\font\tenrm=cmr10
\font\tenit=cmti10 
\font\tenbf=cmbx10
\font\bfit=cmbxti10 at 10pt
\font\ninerm=cmr9
\font\nineit=cmti9

\font\eightrm=cmr8

\def\qed{\hbox{${\vcenter{\vbox{			
   \hrule height 0.4pt\hbox{\vrule width 0.4pt height 6pt
   \kern5pt\vrule width 0.4pt}\hrule height 0.4pt}}}$}}

\begin{document}

\runninghead{Recent Results Using the Overlap Dirac Operator}
{Recent Results Using the Overlap Dirac Operator}

\normalsize\textlineskip
\thispagestyle{empty}
\setcounter{page}{1}

\copyrightheading{}			

\vspace*{0.88truein}

\fpage{1}
\centerline{\bf RECENT RESULTS USING THE OVERLAP DIRAC OPERATOR}
\vspace*{0.37truein}
\centerline{\footnotesize RAJAMANI NARAYANAN}
\vspace*{0.015truein}
\centerline{\footnotesize\it The American Physical Society, One Research
Road, Ridge NY 11961}
\vspace*{0.225truein}

\vspace*{0.21truein}
\abstracts{I derive the overlap
Dirac operator starting from the overlap formalism,
discuss the numerical hurdles in dealing with
this operator and present ways to overcome them.}
{}{}


\vspace*{1pt}\textlineskip	
\section{Introduction}	
\vspace*{-0.5pt}
\noindent

The overlap Dirac operator\cite{n1}, derived from the
overlap formalism\cite{nn1} for the special case of vector
gauge theories, is a way to realize exact chiral symmetry on
the lattice. Exact chiral symmetry on the lattice does come at a price --
numerical implementation of the overlap Dirac operator
is significantly more expensive than Wilson or staggered
operator. In spite of this numerical hurdle,
we already have several physics results in quenched gauge
theories using the overlap Dirac operator:
\begin{romanlist}
\item 
Evidence for spontaneous chiral symmetry breaking at zero
temperature\cite{ehn2}.
\item 
Evidence for chiral symmetry breaking in the deconfined
phase possibly due to a dilute gas of 
instanton and anti-instantons\cite{ehkn1}.
\item Evidence for a diverging chiral
condensate in the two dimensional U(1) case\cite{kn1}.
\item A study of exact zero modes of
overlap fermions in the adjoint representation
lend some support to the existence of
fractional topological charge\cite{ehn3}.
\end{romanlist}
In this talk, I shall derive the overlap
Dirac operator starting from the overlap formalism,
discuss the numerical hurdles in dealing with
this operator and present ways to overcome them.

\section{ The Overlap formalism}

The determinant of the chiral Dirac Operator
$C = \sigma_\mu ( \partial_\mu + i A_\mu ) $
can be realized on the lattice as an overlap of 
two many body states\cite{nn2,nn1},
namely;
\begin{equation}
 \det C = < 0 - | 0 + >,
\end{equation}
where $|0\pm>$ are many body ground states of $a^\dagger H(m) a$
and $a^\dagger \gamma_5 a$ respectively. 
The $a^\dagger$ and $a$ are canonical fermion creation and destruction
operators and
$\gamma_5H(m)$ is a massive Dirac operator on the lattice
with the mass set to a value less than zero. 
One choice is the Wilson Dirac operator, $H(m)=H_w(m)$. 
This realization of the chiral Dirac operator is {\it natural}
since
$C$ is an operator that maps two different spaces, namely spinors
under the (0,1/2) representation to (1/2,0) representation.
Therefore $C$ does not have an eigenvalue problem and
the determinant of $C$ is a map between the highest
form in the two spaces 
connected by the operator $C$.

Clearly, the overlap formula does not fix the phase of $|0+\rangle$ since
it is only defined as an eigenvector of a Hamiltonian and this
is how it should be since the chiral determinant is a map between
two different spaces.
The details
involved in the phase choice and possible gauge breaking is the
subject of chiral gauge theories. 
For vector gauge theories,
we want
$\det CC^\dagger = | \langle 0 - | 0 + \rangle |^2$
and the phase choice does not matter indicating a trivial cancelation
of anomalies.

Computing the overlap of two many body states seems like a insurmountable
numerical task in four dimensional theories since one has to
diagonalize $H_w$, form
the many body state from the negative energy single particle states
and compute the overlap by computing 
a determinant of a dense matrix, half the size of $H_w$.
But there is an elegant solution to circumvent these steps by
directly dealing with the many body states and this is the overlap
Dirac operator\cite{n1}.

The massless overlap Dirac operator 
\begin{equation}
D_o={1\over 2} \Bigl [ 1 + \gamma_5\epsilon(H_w) \Bigr ]
\end{equation}
is derived from the overlap formalism
as follows. Let $U$ be the unitary matrix that diagonalizes $H_w$:
\begin{equation}
H_w U = U \lambda;\ \ \ \  U = \pmatrix { \alpha & \gamma \cr \beta & 
\delta \cr };\ \ \ \ 
\epsilon(H_w)\pmatrix { \alpha & \gamma \cr \beta & 
\delta \cr } = \pmatrix { \alpha & - \gamma \cr \beta & 
- \delta \cr } 
\end{equation}
Using
$ \det U = \det \alpha / \det \delta^\dagger $,
we derive
 \begin{eqnarray} 
| \langle 0 - | 0 + \rangle |^2  & = & \det \delta \det \delta^\dagger 
= \det \delta \det \alpha \det U^\dagger 
=  \det \pmatrix{ \alpha & 0 \cr 0 & \delta \cr} U^\dagger \nonumber \\
& = & \det {1\over 2} \Biggl\{ 
\pmatrix { \alpha & \gamma \cr \beta & \delta \cr}
+ \pmatrix { \alpha & -\gamma \cr -\beta & \delta \cr} \Biggr\} U^\dagger 
\nonumber\\
& = & \det {1\over 2} \Biggl\{ 
\pmatrix { \alpha & \gamma \cr \beta & \delta \cr}
+ \pmatrix{ 1 & 0 \cr 0 & -1 \cr }
\pmatrix { \alpha & -\gamma \cr \beta & -\delta \cr} \Biggr\} U^\dagger 
\nonumber \\
& = & \det {1\over 2} \Bigl [ U +\gamma_5 \epsilon(H_w) U \Bigr ] U^\dagger 
= \det {1\over 2} \Bigl [ 1 + \gamma_5\epsilon(H_w) \Bigr ] 
\end{eqnarray}

It is not immediately clear as to how
it helps numerically since one will have to deal with $\epsilon(H_w)$
without having to diagonalize $H_w$.  
There are two possible approaches.
One approach is to use Gegenbauer 
polynomials to represent $\sqrt{H_w^2}$\cite{b1,hjl1}.
Typically one need to go to
a high order polynomial and this method is not expected to
efficient.
The other approach is to use
the rational approximation\cite{n2}
where one approximates $\epsilon(H_w)$ as a sum of poles
\begin{equation}
\epsilon(H_w) = c_0 + \sum_{i=1}^n {c_i H_w \over H_w^2 + d_i}
\end{equation}
Using the
method of multiple masses, one action of $\epsilon(H_w)$ on a
vector can be realized by a single conjugate gradient algorithm
independent of the number of poles. 
This makes it numerically
quite attractive.

\section { Spectrum of the quenched $H_w$ }

$\epsilon(H_w)$ is discontinuous at the zero of $H_w$. 
Approximations have to be good up to lowest eigenvalue of $H_w$
and this can be a problem if $H_w$ has very small eigenvalues.
The density of the spectrum of $H_w(m)$, $\rho(\lambda)$, in
a quenched ensemble has a non-zero $\rho(0)$ at any fixed lattice
coupling at the values of $m$ that are relevant ($m < m_c$)\cite{ehn1}.
This numerical result has support from an analytical argument
where one shows that small defects can already
give rise to a gapless spectrum\cite{bnn1}.

One can also show that
a change in gauge field topology necessitates zero eigenvalues at any mass.
To see this,
let us assume we have a gauge field configuration that
has zero topology. Then $H_w(m)$ has an equal number of + and -
eigenvalues. 
Consider evolving from this configuration 
to another gauge field
configuration that has non-zero topology. This configuration
has a spectrum where the number of + and - eigenvalues of $H_w(m)$
are not equal. 
The spectrum as a function of the evolution has one configuration in
the path where $H_w(m)$ has a zero eigenvalue. In a discrete evolution
scheme the exact zero will be avoided but one can have arbitrarily
small eigenvalues.

Therefore one will have to live with very small
eigenvalues of $H_w(m)$ or its variants.
Numerical techniques that deal with
$\epsilon(H_w)$ will have to project out a few small eigenvectors and treat
them exactly. On a finite lattice and at a fixed lattice spacing,
the number of eigenvalues below a fixed number $\lambda_{\rm min}$ will
grow with volume since $\rho(0)$ is finite. This would mean that one
has to project out more eigenvalues as one increases the volume
and/or go to a larger number of poles in the rational approximation.
It is useful to compare the overlap formalism
with the related method used to realize
chiral symmetry on the lattice, namely domain wall fermions\cite{d1}. This
is a five dimensional realization 
and the effective overlap Dirac operator is
obtained by setting $H=H_d=\log(T_w)$ where $T_w$ is the transfer matrix
in the fifth direction\cite{n3}. The low lying spectrum of $H_d$ is
completely governed by the low lying spectrum of $H_w$ and hence
the problems caused by a finite $\rho(0)$ exist for domain wall
fermions\cite{ehn4}. In practice one works with a finite extent in the fifth
direction ($L_s$) and this amounts to an approximation of the $\epsilon(H_w)$
by $\tanh({1\over 2}L_sH_d)$. Clearly, small eigenvalues are not taken care
of properly at a finite $L_s$ and one will have to go to larger
$L_s$ as one increases the lattice volume at a fixed lattice spacing.
Current simulations using domain wall fermions\cite{ba1} seem to indicate
a significant effect due to finite $L_s$. One can avoid this
by projecting out small eigenvalues and treating them exactly in
the domain wall formalism\cite{eh1}.

Each action of $\epsilon(H_w)$ requires
a Conjugate Gradient type algorithm and therefore the solution
to the equation of the form
$D_o(m) \psi = b$
requires nested Conjugate Gradient.
Is it numerically much more involved than
domain wall fermions since it only involves one inversion
of a higher dimensional operator?
One can write down a five dimensional operator from which
one gets the required four dimensional overlap Dirac operator by
integrating out all but one fermion degree of freedom.
An analysis of the condition numbers shows that the the
five dimensional inversion is no less expensive that two nested
conjugate gradients\cite{nn3}.
In the nested case, it is easy to see
that the condition number is proportional to the product of the
condition number
of $H_w$ and the fermion mass, $\mu$. 
This also turns out to
be the case for the five dimensional version and for the
conventional domain wall fermions. 
This shows that it is practical to work directly with the four
dimensional operator.

\nonumsection{Acknowledgements}
\noindent
I would like to thank Robert Edwards, Urs Heller, Joe Kiskis and
Herbert Neuberger for many useful discussions.

\nonumsection{References}

\end{document}